\documentclass{svproc}
\usepackage{graphicx}
\begin{document}
\mainmatter              
\title{Primordial black holes as dark matter
and generators of cosmic structure}
\titlerunning{Primordial black holes}  
%
\author{Bernard Carr}
%
\authorrunning{Carr}
%
\tocauthor{Carr}
\institute{School of Physics and Astronomy, Queen Mary University of London \\
\email{B.J.Carr@qmul.ac.uk}
}
\maketitle  

\begin{abstract}
Primordial black holes (PBHs) could provide the dark matter but a variety of 
constraints restrict the possible mass windows to $10^{16} - 10^{17}$g, $10^{20} - 10^{24}$g and $10 - 10^3M_{\odot}$. The last possibility is of special interest in view of the recent detection of black-hole mergers by LIGO.  PBHs larger than $10^3 M_{\odot}$ might have important cosmological consequences even if they have only a small fraction of the dark matter density. In particular, they could generate cosmological structures either individually through the `seed' effect or collectively through the `Poisson' effect, thereby alleviating some problems associated with the standard cold dark matter scenario.
\end{abstract}
\section{Introduction}

Primordial black holes (PBHs) have been a source of  interest for nearly 50 years \cite{ZeldovichNovikov69}, despite the fact that there is still no evidence for them. One reason for this interest is that only PBHs could be small enough for Hawking radiation to be important \cite{Hawking:1974rv}. This has not yet been confirmed experimentally and there remain major conceptual puzzles associated with the process.
Nevertheless, this discovery is generally recognised as one of the key developments in 20th century physics because it beautifully unifies general relativity, quantum mechanics and thermodynamics. The fact that Hawking was only led to this discovery through contemplating the properties of PBHs illustrates that it can be useful to study something even if it does not exist! But, of course, the situation is much more interesting if PBHs do exist.

PBHs smaller than about $10^{15}$g would have evaporated by now with many interesting cosmological consequences \cite{Carr:2009jm}. Studies of such consequences have placed useful constraints on models of the early Universe and, more positively, evaporating PBHs have been invoked to explain certain features: for example, the extragalactic \cite{Page:1976wx,Carr:1976jy} and Galactic \cite{1996ApJ...459..487W,Lehoucq:2009ge} $\gamma$-ray backgrounds, antimatter in cosmic rays \cite{Kiraly:1981ci,MacGibbon:1991vc},
the annihilation line radiation from the Galactic centre \cite{1980A&A....81..263O,Bambi:2008kx}, the reionisation of the pregalactic medium \cite{1997grgp.conf..413G,Belotsky:2014twa} and some short-period $\gamma$-ray bursts \cite{1996MNRAS.283..626B,Cline:1996zg}. 
However,  there are usually other possible explanations for these features, so there is no definitive evidence for evaporating PBHs. Only the original papers for each topic are cited here and a more comprehensive list of references can be found in Ref.~\cite{Carr:2009jm}. 

Attention has therefore shifted to the PBHs larger than $10^{15}$g, which are unaffected by Hawking radiation. Such PBHs might have various astrophysical consequences, such as providing seeds for the supermassive black holes in galactic nuclei \cite{1984MNRAS.206..801C,Bean:2002kx}, the generation of large-scale structure through Poisson fluctuations \cite{Meszaros:1975ef,Afshordi:2003zb} and important effects on the thermal and ionisation history of the Universe \cite{1981MNRAS.194..639C,Ricotti:2007au}.
Again only the original papers are cited here. But perhaps the most exciting possibility
is that they could provide the dark matter which comprises $25\%$ of the critical density \cite{Frampton:2015xza,Carr:2016drx}, an idea that goes back to the earliest days of PBH research \cite{1975Natur.253..251C}. Since PBHs formed in the radiation-dominated era, they are not subject to the well-known cosmological nucleosynthesis constraint that baryons can have at most $5\%$ of the critical density \cite{Cyburt:2003fe}. They should therefore be classed as non-baryonic and 
behave like any other form of cold dark matter (CDM).

As with other CDM candidates. there is still no compelling evidence that PBHs provide the dark matter. 
There have been claims that the  microlensing of quasars could indicate dark matter in  jupiter-mass PBHs  \cite{1993Natur.366..242H} but these are controversial. There was also a flurry of excitement about PBHs  in 1997, when the 
MACHO microlensing results \cite{Alcock:1996yv} suggested that the dark matter could be in compact objects of mass $0.5\,M_{\odot}$. Alternative dark matter candidates could be excluded and PBHs of this mass might naturally form at the quark-hadron phase transition at $10^{-5}$s \cite{Jedamzik:1998hc}. Subsequently, however, it was shown that such objects could comprise only $20\%$ of the dark matter and indeed the entire mass range $10^{-7}\,M_{\odot}$ to $10\,M_{\odot}$ is now excluded from providing the dark matter \cite{Tisserand:2006zx}. 

In recent decades attention has focused on other mass ranges in which PBHs could provide the dark matter and  numerous constraints 
allow only three possibilities:  the asteroid mass range  ($10^{16}$ -- $10^{17}$g),
the sublunar mass range ($10^{20}$ -- $10^{26}$g) and 
the intermediate mass black hole (IMBH) range ($10$ -- $10^{3}\,M_{\odot}$).  There is particular interest in the last possibility because the coalescing black holes detected by LIGO \cite{Abbott:2016blz} could be of primordial origin, although this would not necessarily require the PBHs to provide all the dark matter. Also PBHs could have important cosmological consequences even if they provide only a small fraction of the dark matter,
 so we explore this possibility below.

\section{PBH formation}

PBHs could have been produced during the early Universe due to various mechanisms. Matching the cosmological density at a time $t$ are the big bang with density required to form a PBH of mass $M$
implies that the PBH mass is comparable to the horizon mass at formation \cite{Hawking:1971ei,Carr:1974nx}:
\begin{equation}
	M
		\sim
								\frac{ c^{3}\, t }{ G }
		\sim
								10^{15}
								\left(
									\frac{t}{10^{-23}\rm{s}}
								\right)\,
								\rm{g}
								\, .
								\label{eq:Moft}
\end{equation}
Hence PBHs could span an enormous mass range: those formed at the Planck time ($10^{-43}$s) would have the Planck mass ($10^{-5}$g), whereas those formed at $1$~s would be as large as $10^{5}\, M_{\odot}$.
By contrast, black holes forming at the present epoch (eg. in the final stages of stellar evolution) could never be smaller than about $1\,M_{\odot}$. In some circumstances PBHs may form over an extended period, corresponding to a wide range of masses, but their spectrum could be extended even if they form at a single epoch. 

As discussed in numerous papers, starting in the 1990s~\cite{Carr:1994ar,Ivanov:1994pa,Randall:1995dj,GarciaBellido:1996qt}, the quantum fluctuations arising in various inflationary scenarios are a possible source of PBHs. In some of these scenarios the fluctuations generated by inflation are ``blue'' (i.e.~decrease with increasing scale) and this means that the PBHs form shortly after reheating.
Others involve some form of ``designer'' inflation, in which the power spectrum of the fluctuations{\,---\,}and hence PBH production{\,---\,}peaks on some scale.
In other scenarios, the fluctuations have a ``running index'', so that the amplitude increases on smaller scales but not according to a simple power law.
PBH formation may also occur due to some sort of parametric resonance effect before reheating, 
in which case the fluctuations tend to peak on a scale associated with reheating. This is usually very small but several scenarios involve a secondary inflationary phase which boosts this scale into the macroscopic domain. There are too many papers on these topics to cite here but a comprehensive list of
references can be found in Ref.~\cite{Carr:2009jm}. 

Whatever the source of the inhomogeneities, PBH formation would be enhanced if there was a  reduction in the pressure at some epoch - for example, at the QCD era \cite{Crawford:1982yz,Byrnes:2018clq} 
or if the early Universe went through a dust-like phase  as a result of being dominated by non-relativistic particles for a period \cite{1982SvA....26..391P,Carr:2017edp} 
or undergoing slow reheating after inflation \cite{Khlopov:1985jw,Carr:2018nkm}. 
Another possibility is that PBHs might have formed spontaneously at some sort of phase transition, even if there were no prior inhomogeneities, for example from the collisions of bubbles of broken symmetry \cite{Kodama:1982sf,Hawking:1982ga}
or the collapse of cosmic strings \cite{Hawking:1987bn,Polnarev:1988dh}
or domain walls \cite{Rubin:2001yw,Dokuchaev:2004kr}.
Further references for such  models can be found in Ref.~\cite{Carr:2009jm}. 

The fraction of the mass of the Universe in PBHs is time-dependent but its value at the PBH formation epoch is of particular interest. If the PBHs formed at a redshift $z$ or time $t$ and contribute a fraction  $f(M)$ of the dark matter on some mass scale $M$, then the collapse fraction is \cite{Carr:1975qj}
\begin{equation}
	\beta (M) = f(M) \left( \frac{1+z}{1+z_{eq}} \right) \sim
								10^{-6} f(M) \left( \frac{t}{1 \, \rm{s}} \right)^{1/2}
		\sim
								10^{-18} f(M) \left( \frac{M}{10^{15}\rm{g}} \right)^{1/2}
								\, ,
								\label{eq:roughomega}
\end{equation}
where we assume the PBHs form in the radiation-dominated era, $z_{eq} \approx 4000$ is the redshift of matter-radiation equality,
and we use Eq.~(\ref{eq:Moft}) at the last step. The $(1 + z)$ factor arises because the radiation density scales as $(1 + z)^{4}$\,, whereas the PBH density scales as $(1 + z)^{3}$. Any limit on $f(M)$ (eg. $f \leq 1$ for $M > 10^{15}$g) therefore places a constraint on $\beta( M )$, which is necessarily tiny.  

On the other hand, one also {\it expects} the collapse fraction to be small. For example, if the PBHs form from primordial inhomogeneities which are Gaussian with rms amplitude $\delta_H(M)$ at the horizon epoch, one predicts  \cite{Carr:1975qj}
\begin{equation}
\beta (M) \approx \rm{erfc} \left( \frac{\delta_c}{ \delta_H(M)} \right) \, ,
\end{equation}
where `erfc' is the complimentary error function and $\delta_c  \approx 0.4$ is the threshold for collapse against the pressure \cite{Musco:2004ak,Harada:2013epa}. In a dust era, the collapse fraction is $\beta \sim 0.02 \delta_H(M)^5$, corresponding to the probability of sufficient spherical symmetry, but this is still small \cite{Khlopov:1980mg,Harada:2017fjm}. In the other scenarios, $\beta$ depends upon some cosmological parameter (eg. the string tension or bubble formation rate).

\section{Constraints on 
non-evaporated black holes}

\begin{figure}
	\begin{center}
\includegraphics[scale=.55]{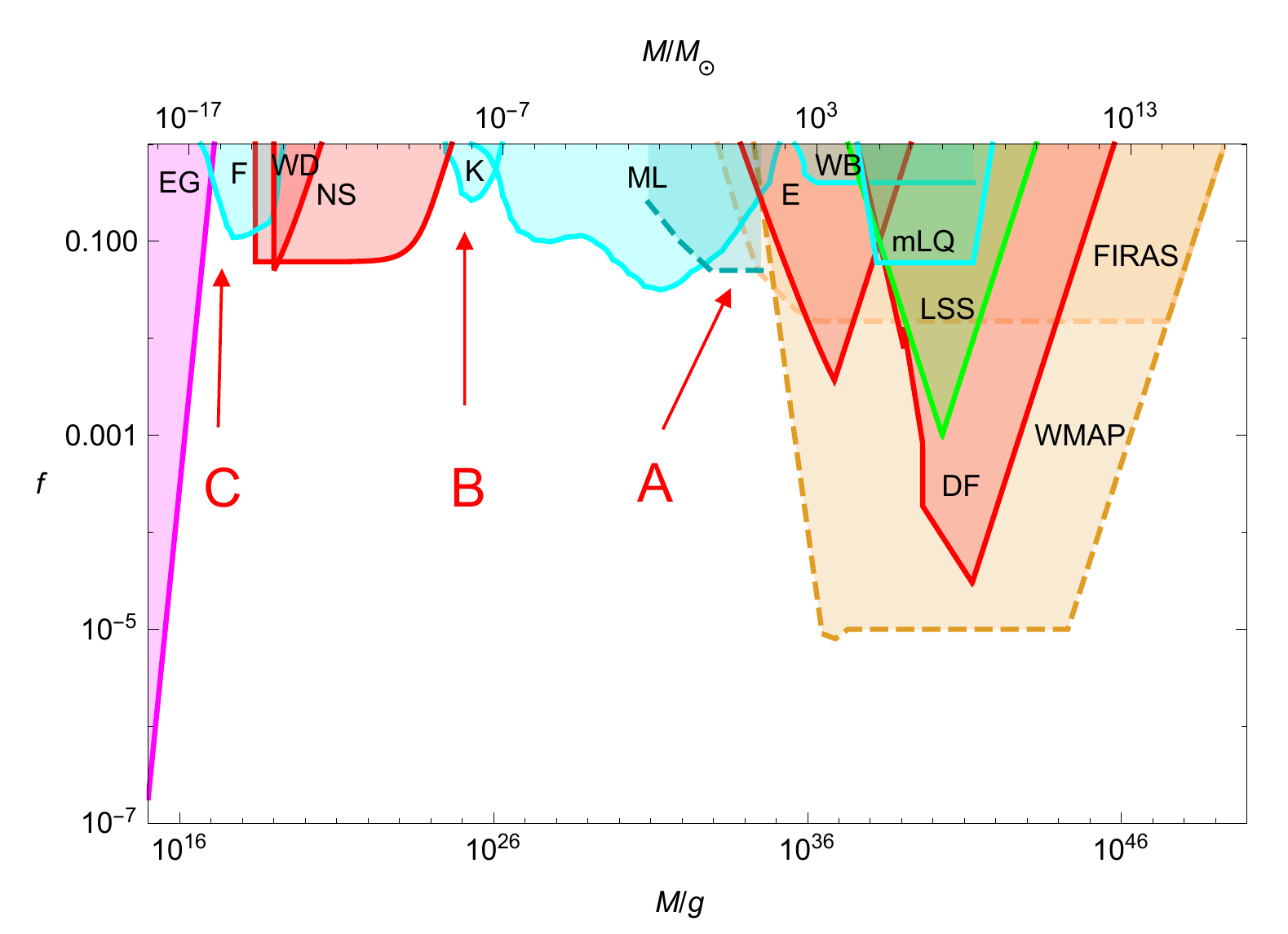}
	\end{center}
	\caption{
		Constraints on $f( M )$  from Ref.~\cite{Carr:2016drx} for a monochromatic mass function from a variety of evaporation (magenta), dynamical (red), lensing (cyan), 
		large-scale structure (green) and accretion (orange) effects.  
See text for references and more recent limits.
		 The accretion limit is shown with a broken line since this
has now been improved. 
The arrows indicate the three mass windows where $f$ can be close to $1$. } 	
	\label{fig:large}
\end{figure}

The constraints on $f( M )$,  the fraction of the halo in PBHs of mass $M$,   are summarised in Fig.~\ref{fig:large}, which is taken from Ref.~\cite{Carr:2016drx}, although some of them have now been revised. 
All the limits assume that the PBHs have a monochromatic mass function and cluster in the Galactic halo in the same way as other forms of CDM. 
The effects are extragalactic $\gamma$-rays from evaporation (EG) \cite{Carr:2009jm},
		femtolensing of $\gamma$-ray bursts (F) \cite{Barnacka:2012bm}, 
		white-dwarf explosions (WD) \cite{Graham:2015apa},
		neutron-star capture (NS) \cite{Capela:2013yf}, 
		Kepler microlensing of stars (K) \cite{Griest:2013aaa}, 
		MACHO/EROS/OGLE microlensing of stars (ML) \cite{Tisserand:2006zx}
		and quasar microlensing
(ML) \cite{Mediavilla:2009um}, 
		survival of a star cluster in Eridanus II (E) \cite{Brandt:2016aco}, 
		wide-binary disruption (WB) \cite{Quinn:2009zg}, 
		dynamical friction on halo objects (DF) \cite{Carr:1997cn}, 
		millilensing of quasars (mLQ) \cite{Wilkinson:2001vv}, 
		generation of large-scale structure through Poisson fluctuations (LSS) \cite{Afshordi:2003zb}, 
		and accretion effects (WMAP, FIRAS) \cite{Ricotti:2007au}. 

As indicated by the arrows in Fig.~\ref{fig:large} , the permittted  mass windows for $f \sim 1$ are: (A) the intermediate mass range ($10$ -- $10^{3}\,M_{\odot}$); (B) the sublunar mass range ($10^{20}$ -- $10^{24}$g); and  (C)
the asteroid mass range  ($10^{16}$ -- $10^{17}$g). However, there are further limits since Fig.~\ref{fig:large} was produced and some people claim that  even these windows are now excluded. For example, scenario C  may be ruled out by observations of the Galactic $\gamma$-ray background \cite{Carr:2016hva} or positron flux \cite{Boudaud:2018hqb}. One problem with scenario A is that such objects would disrupt wide binaries in the Galactic disc. It was originally claimed that this would exclude objects above $400\,M_{\odot}$
but more recent studies may reduce this mass \cite{Monroy-Rodriguez:2014ula}, so the narrow window between the microlensing and wide-binary bounds is shrinking. There are new microlensing constraints in the lunar-mass range from the Subaru telescope \cite{Niikura:2017zjd} and in the intermediate mass  range from supernovae \cite{Zumalacarregui:2017qqd}. Also the CMB accretion constraints have been revised and are now weaker \cite{Ali-Haimoud:2016mbv}, although there are new accretion limits from X-ray observations \cite{Poulin:2017bwe,Inoue:2017csr}.  Two talks at this symposum imply  interesting new constraints associated with tidal streams \cite{Bovy:2016irg} and lensing substructure \cite{Hezaveh:2016ltk}.

The PBHs in either scenario A and B could be generated by inflation but theorists are split as to which window they favour. For example, Inomata et al. \cite{inomata:2017okj} argue that doube inflation can produce a peak at around $10^{20}$g, while Clesse and Garcia-Beillido \cite{Clesse:2015wea} argue that hybrid inflation can produce a peak at around $10 M_{\odot}$, this being favored by the LIGO results.
A peak at this mass could also be produced  by a reduction in the pressure at the quark-hadron phase transition \cite{Byrnes:2018clq}, even if the primordial fluctuations have no feature on that scale. 
There is a parallel here with the search for particle dark matter,
where there is also a split between  groups searching for light and heavy candidates.

The constraints discussed above assume that the PBH mass function is monochromatic (i.e. 
with a width $\Delta M \sim M$). However, 
there are many scenarios 
in which one would expect the mass function to be extended. For example. inflation tends to produce a lognormal mass function \cite{Dolgov:1992pu} and critical collapse generates an extended low mass tail  \cite{Yokoyama:1998xd,Musco:2012au}. In the context of the dark-matter problem, this is a two-edged sword \cite{Carr:2016drx}. On the one hand, it means that the {\it total} PBH density may suffice to explain the dark matter, even if the density in any particular mass band is small and within the observational bounds discussed above. On the other hand, even if PBHs can provide all the dark matter at some mass-scale, the extended mass function may still violate the constraints at some other scale. 
This issue been addressed in a number of recent papers \cite{Green:2016xgy,Carr:2017jsz,Kuhnel:2017pwq}, though with somewhat different conclusions.

\section{Effects of PBHs on cosmic structures}

PBHs of mass $m$ provide a source of fluctuations for objects of mass $M$ in two ways: (1) via the seed effect, in which the Coulomb effect of a {\it single} black hole generates an initial density fluctuation $m/M$;
 (2) via the Poisson effect, in which the $\sqrt{N}$ fluctuation in the number of black holes generates an initial density fluctuation $(fm/M)^{1/2}$ for a PBH dark matter fraction $f$. Both types of fluctuations then grow through gravitational instability to bind  regions of mass $M$.  
Each of these proposals has a long history and detailed references can be found in Ref.~\cite{Carr:2018rid}.  
The relationship between the two mechanisms is  subtle, so we will consider both of them below
and determine the dominant one for each mass scale. 

\begin{figure}
 \begin{center}
\includegraphics[scale=.35]{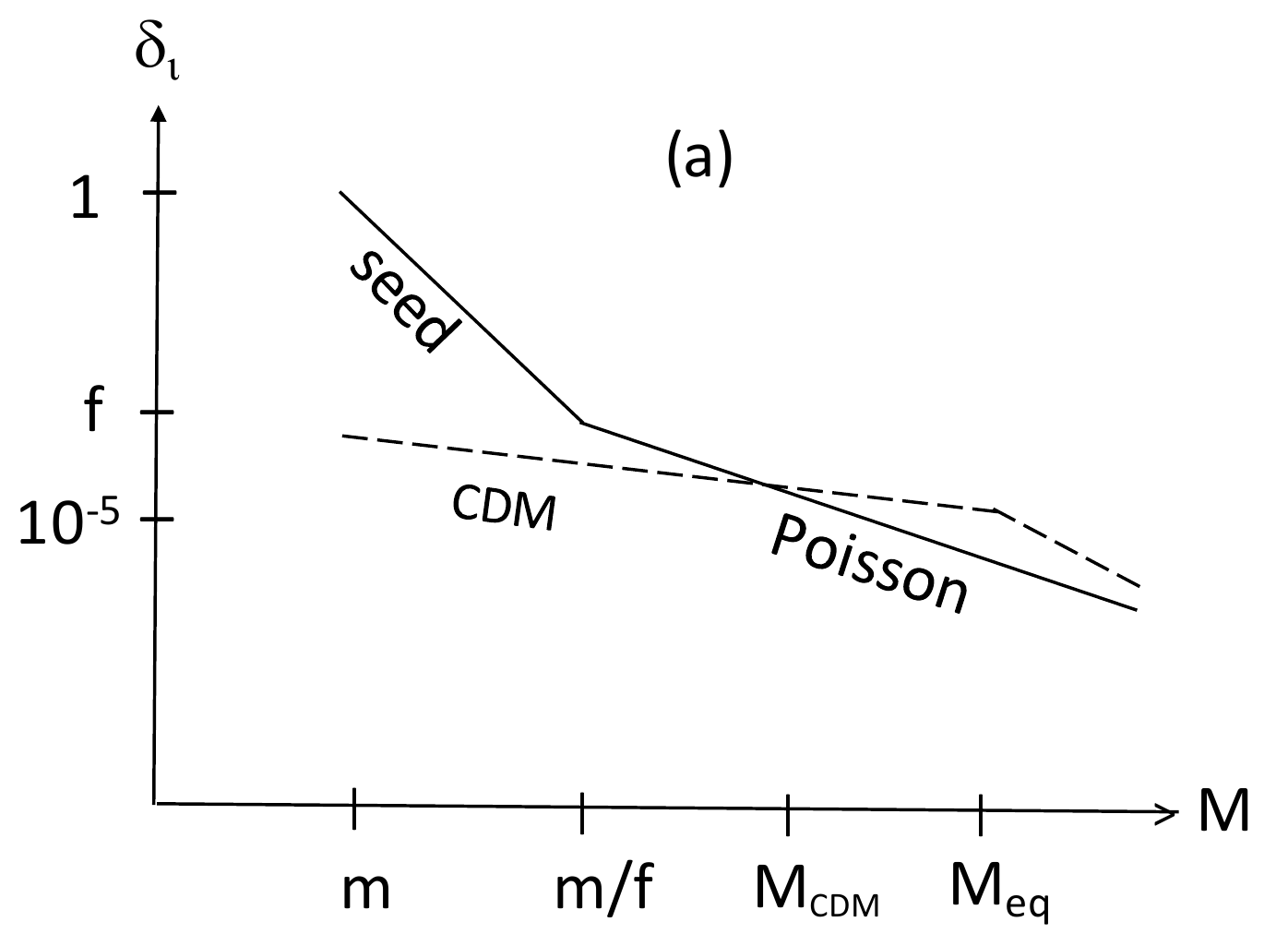}
\includegraphics[scale=.35]{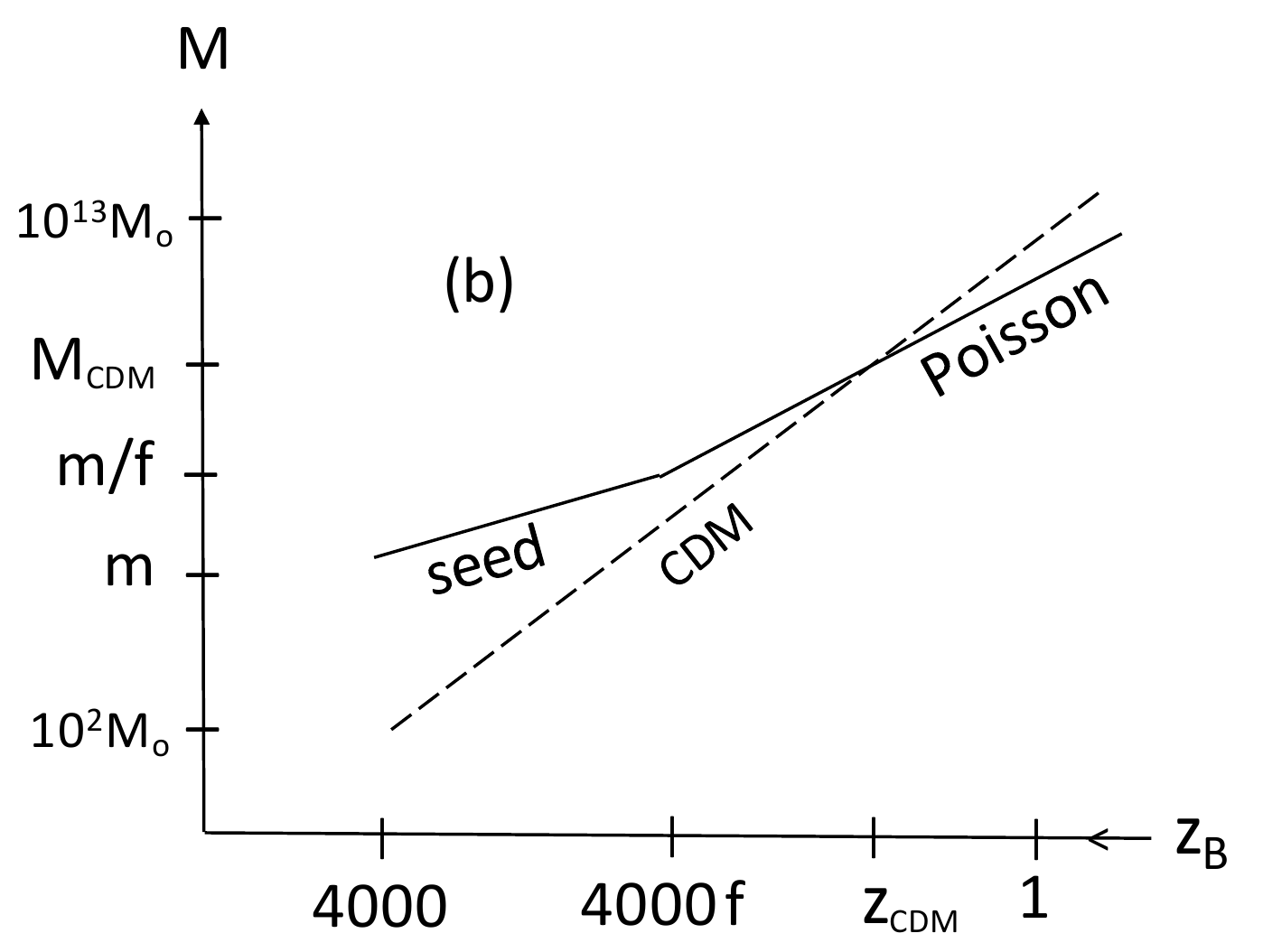}
  \caption {(a) Form of initial fluctuation $\delta_i$ as a function of $M$ for the seed and Poisson effect with fixed $f$, the first dominating 
at small $M$ if $f $ is small but the second always dominating if $f \sim 1$. 
(b) Mass $M$ binding at redshift $z_B$ for fixed $f$, the Poisson effect dominating for low $z$ if $f$ is small  but at all $z$ if $f \sim 1$.
Also shown by dashed lines  are the forms for $\delta_i$ and $M(z_B)$ predicted by the CDM model, this indicating  the range $M > M_{CDM}$ and $z_B < z_{CDM}$ for which CDM fluctuations dominate. From Ref.~\cite{Carr:2018rid}.}
 \label{fig2}
 \end{center}
 \end{figure}

If the PBHs have a single mass $m$, 
the initial fluctuation in the matter density on a scale $M$ is
\begin{equation}
\delta_i \approx
m/M
\,  (\mathrm{seed}) \, ,        
\quad (f m/M)^{1/2}
\, (\mathrm{Poisson}) \, ,
\label{initial}
 \end{equation}
where 
$M$ excludes the radiation content. If PBHs provide the dark matter, $f \sim 1$ and the Poisson effect dominates for all $M$ but we also consider scenarios with $f \ll 1$. The Poisson effect then dominates for $M > m/f$ and the seed effect for $M < m/f$. Indeed, the first expression in (\ref{initial}) {\it only} applies  for $f < m/M$,
since otherwise a region of mass $M$ would be expected to contain more than one black hole.
The dependence of $\delta_i$ on $M$ is indicated in Fig.~\ref{fig2}(a).
The fluctuation grows as $(1+z)^{-1}$ from the redshift of matter-radiation equality, 
$z_{eq} \approx 4000$, until it binds when $\delta \approx 1$. Therefore the mass binding at redshift $z_B$ is 
\begin{equation}
M \approx
4000 \, m z_B^{-1} \, 
 (\mathrm{seed}) \, ,  \quad     
10^7 f m z_B^{-2}
\, (\mathrm{Poisson}) \, ,
\label{bind}
 \end{equation}
as illustrated in Fig.~\ref{fig2}(b). The CDM fluctuations are shown for comparison. These always dominate at sufficiently large scales but the PBHs provide an extra peak in the power spectrum on small scales.

One can place interesting upper limits on $f(m)$
by requiring that various types of structure do not form too early~\cite{Carr:2018rid}. One can also take a more positive approach, exploring the possibility that PBHs may have {\it helped} the formation of these objects, thereby complementing the standard CDM scenario of structure formation.
If the PBHs have a monochromatic mass function and provide {\it all} the dark matter ($f \sim 1$), then the Poisson effect dominates on all scales and various astrophysical constraints discussed above require $m < 10^2M_{\odot}$. This implies that PBHs can only bind subgalactic masses
but still  allows them to play a role in producing the first bound baryonic clouds
or the SMBHs which power quasars.

For $f \ll 1$, the seed effect dominates on small scales and can bind a region of up to $4000$ times the PBH mass. 
It is known that most galaxies contain central supermassive black holes with a mass proportional  to the bulge mass \cite{2015ApJ...813...82R}
and this correlation is naturally explained by the seed effect if the black holes are primordial - rather than forming after galaxies - with an extended mass function. However, limits on
 the $\mu$-distortion in the CMB due to the dissipation of fluctuations before decoupling exclude many PBHs larger than $10^5 M_{\odot}$ unless one invokes non-Gaussian fluctuations or accretion \cite{Nakama:2017xvq}.

\section{LIGO gravitational-wave limits}

The proposal that the dark matter could comprise PBHs in the intermediate mass range has attracted much attention recently as a result of the LIGO detections of merging binary black holes with mass around $30\,M_{\odot}$  \cite{TheLIGOScientific:2016pea,TheLIGOScientific:2016wfe,Abbott:2016nmj,LIGOScientific:2018jsj}. Since the black holes are larger than initially expected, it has been suggested that they could represent a new  population.
One possibility is that they were of Population III origin (i.e. forming between decoupling and galaxies).  The suggestion that LIGO might detect gravitational waves from coalescing  intermediate mass  Population III black holes was first made 
more than 30 years ago \cite{1984MNRAS.207..585B} and - rather remarkably - Kinugawa et al. predicted a Population III coalescence peak at $30 M_{\odot}$ shortly before the first LIGO detection \cite{Kinugawa:2014zha}.

Another possibility - more relevant to the present considerations 
- is that the LIGO black holes were primordial, as first discussed  in Ref. \cite{Nakamura:1997sm}.
 This does not necesarily require the PBHs to provide {\it all} the dark matter. While several authors have made this connection \cite{Bird:2016dcv,Clesse:2016vqa}, the predicted merger rate depends on when the binaries form and 
uncertain astrophysical factors, so others argue that the dark matter fraction could be small
\cite{Sasaki:2016jop,Nakamura:2016hna,Sasaki:2018dmp}.
 Indeed the LIGO results have been used to constrain the PBH dark matter fraction \cite{Raidal:2017mfl,Ali-Haimoud:2017rtz}.
Note that the PBH density should peak at a lower mass than the coalescence signal for an extended PBH mass function, since the gravitational wave amplitude scales as the black hole mass. Indeed, Clesse \& Garcia-Bellido argue that a lognormal distribution  centred at around $3 M_{\odot}$ would naturally explains both the dark matter and the LIGO bursts without violating any of the current PBH constraints \cite{Clesse:2016vqa}.  

A population of massive PBHs would also be expected to generate a stochastic background of gravitational waves \cite{1980A&A....89....6C}, whether or not they form binaries.
If the PBHs have an extended mass function, incorporating both dark matter at the low end and galactic seeds at the high end, this would have important implications for the predicted gravitational wave background. Theorists usually focus on  the gravitational waves  generated by either stellar black holes (detectable by LIGO) or  
supermassive black holes (detectable by  LISA). However, with an extended PBH mass function, the gravitational wave background should encompass both these limits and also every intermediate frequency. 
If the PBHs form from  scalar perturbations of inflatioary origin, there is also a gravitational wave background due to  associated second-order tensor perturbations and this gives very strong potential limits on $f(M)$ \cite{Saito:2009jt,Bugaev:2010bb,Bartolo:2018rku}.

\section{Summary}

In recent years PBHs have been invoked for three purposes:  (1) to explain the dark matter; (2) to provide a source of LIGO coalescences; (3) to
alleviate some of the problems associated with the CDM scenario. In principle, these are distinct roles and any one of them would justify the study of PBHs. On the other hand, if PBHs have an extended mass function, they could play more than one or even all these roles. 

As regards (1), there are only a few mass ranges in which PBHs could provide the dark matter. We have focused particularly on the intermediate mass range $10\,M_{\odot} < M < 10^{3}\,M_{\odot}$, because this may  be relevant to (2), but the sublunar range $10^{20}$ -- $10^{24}\,$g also remains viable. 
The  asteroid range $10^{16}$ -- $10^{17}\,$g is probably the least plausible. 
We have not discussed the possibility that stable Planck-mass relics of PBH evaporations provide the dark matter \cite{MacGibbon:1987my}. This scenario cannot be excluded but it is impossible to test since Planck-scale relics would be undetectable except through their gravitational effects. 

\begin{figure}
 \begin{center}
\includegraphics[scale=.33]{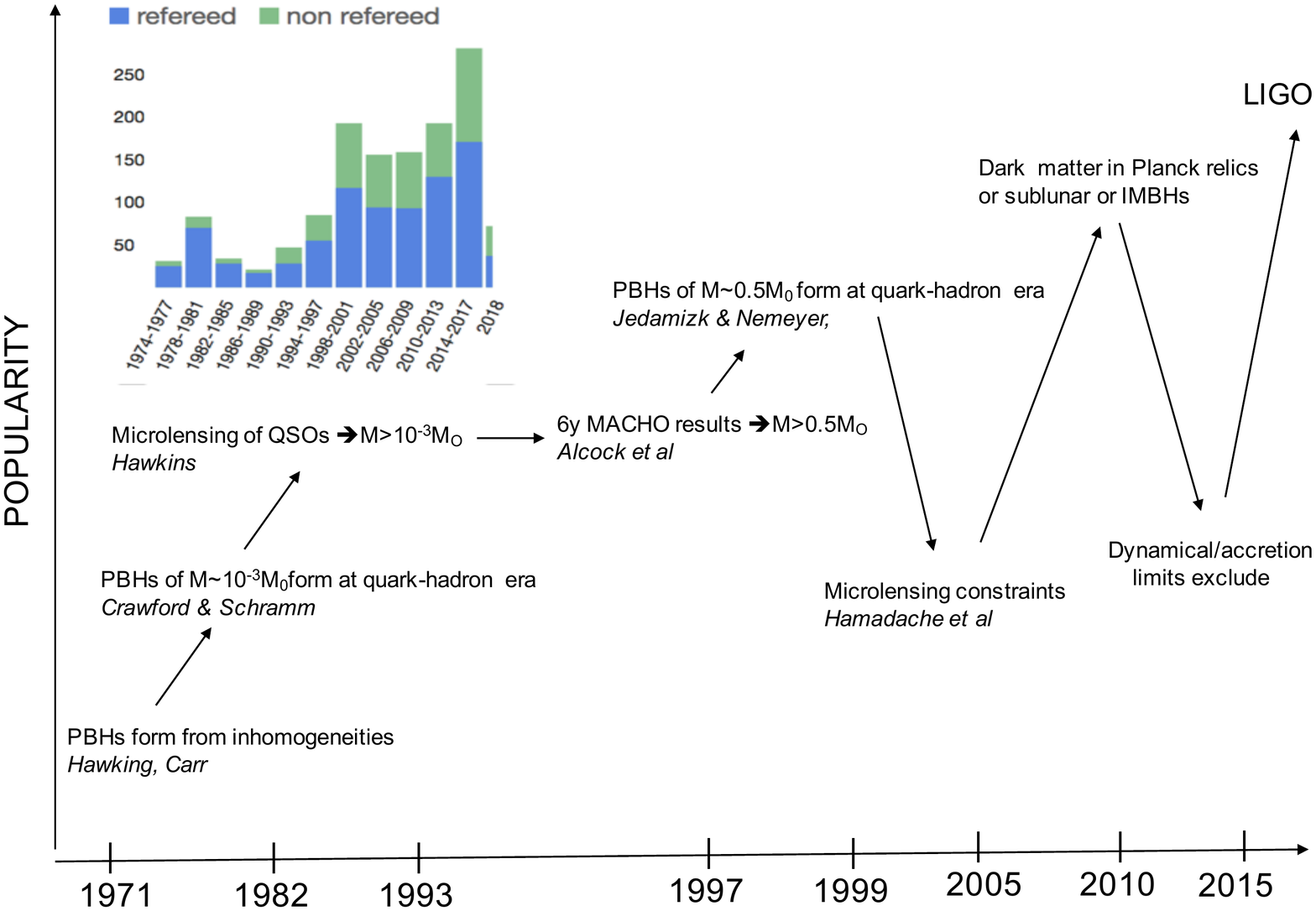}
  \caption {History of popularity of PBHs, as indicated by publication rate (top left).  }
\label{fig3}
 \end{center}
 \end{figure}

Presumably most participants at this meeting would prefer the dark matter to be elementary particles rather than PBHs, so it may be reassuring that for most of the last 50 years the study of PBHs has been a minority interest.
On the other hand, as illustrated in Fig.~\ref{fig3}, PBHs have become increasingly popular in recent years, at least as measured by the annual publication rate on the topic. Indeed, turning to role (3), perhaps the most important point to emphasize, is that PBHs in the intermediate to supermassive mass range could play an important cosmological role even if the do {\it not} provide the dark matter. Perhaps this also
applies for the particle candidates.
Few people would now argue that neutrinos provide the dark matter but they are still extraordinarily important.

\section*{Acknowledgments}
This talk is dedicated to the memory of my friend and mentor Stephen Hawking.  If PBHs turn out to exist,
then his pioneering work on this topic will have been one of his most prescient and important scientific contributions. 
I thank the Simons Foundation for their generous hospitality at this conference and my many PBH coathors over 45 years for an enjoyable collaboration. 

\bibliography{refs}

\clearpage
\bibliographystyle{spphys.bst}

\end{document}